# High-quality 8-fold self-compression of ultrashort near-UV pulses in Ar-filled ultrathin-walled photonic crystal fiber

JIE LUAN,[1,2*] PHILIP ST.J. RUSSELL,[1] AND DAVID NOVOA[3,4,1*]

[1]*Max Planck Institute for the Science of Light, Staudtstrasse 2, 91058 Erlangen, Germany*
[2]*Department of Physics, Friedrich-Alexander-Universität, Staudtstrasse 2, 91058 Erlangen, Germany*
[3]*Department of Communications Engineering, Engineering School of Bilbao, University of the Basque Country (UPV/EHU), Torres Quevedo 1, 48013 Bilbao, Spain*
[4]*IKERBASQUE, Basque Foundation for Science, Plaza Euskadi 5, 48009 Bilbao, Spain*
*jie.luan@mpl.mpg.de & david.novoa@ehu.eus

**Abstract:** We demonstrate generation of 7.6 fs near-UV pulses centered at 400 nm via 8-fold soliton-effect self-compression in an Ar-filled hollow-core kagomé-style photonic crystal fiber with ultrathin core walls. Analytical calculations of the effective compression length and soliton order permit adjustment of the experimental parameters, and numerical modelling of the nonlinear pulse dynamics in the fiber accurately predict the spectro-temporal profiles of the self-compressed pulses. After compensation of phase distortion introduced by the optical elements along the beam path from the fiber to the diagnostics, 71% of the pulse energy was in the main temporal lobe, with peak powers in excess of 0.2 GW. The convenient set-up opens up new opportunities for time-resolved studies in spectroscopy, chemistry and materials science.

## 1. Introduction

Ultrashort pulses in the ultraviolet (UV, 100 to 400 nm) are needed for time-resolved applications in fields such as ultrafast spectroscopy and femtochemistry [1,2], owing to large scattering cross-sections and strong absorption bands that all natural molecules have in this spectral region [3]. There are multiple approaches to generating ultrafast UV light, the most common being non-collinear optical parametric amplifiers [4], which are complex systems that require careful alignment and maintenance, and nonlinear spectral broadening in gas-filled wide-bore capillaries followed by post-compression [5]. Anti-resonant-reflecting hollow-core photonic crystal fibers (ARR-PCFs) [6,7], by permitting low loss guidance in a core some 10 times smaller than wide-bore capillaries, permit operation with few microjoule pulses at much higher (MHz) repetition rates [8]. They also provide enhanced spatial mode quality, excellent broadband UV transparency, a high optical damage threshold, and pressure-tunable dispersion, which taken together allow confinement, manipulation and safe transport of UV light in a precise manner over long distances.

When filled with gas at different pressures, the dispersion and nonlinearity of the ARR-PCFs can be easily adjusted, allowing access to different regimes of operation. In this context, generation of sub-µJ near-UV pulses of ~5.7 fs duration centered at 400 nm was recently demonstrated using soliton-effect self-compression [9]. However, as in other previous studies in, e.g., the near-infrared [10], such short durations could only be indirectly inferred by numerically subtracting the external dispersion introduced by several optical components such as the windows sealing the gas cells, located along the beam path from the fiber endface to the diagnostics. It is worth noticing that most optical materials become increasingly more dispersive at shorter wavelengths, which hampers ultrafast UV applications especially for larger bandwidth pulses. Although removing the windows and working with ambient air partially resolves this issue [11], the compressed pulses are limited in energy, do not reach sub-10-fs duration and undergo significant spectral red-shift as a result of intra-pulse Raman scattering, principally by $N_2$.

Here we report efficient self-compression of ~64 fs-long, chirped near-UV pulses centered at 400 nm (~40 fs bandwidth-limited duration) to ~7.6 fs (compression factor ~8.4) in an ultra-thin-walled, kagomé-style ARR-PCF filled with argon. By judiciously choosing the fiber parameters, namely its length, core size, core-wall thickness, and gas pressure, we could operate with a low soliton order $N \sim 3$, thereby ensuring a fair compromise between compression factor and pulse quality $Q$, which we define as the ratio, in percentage, between the energy contained within the main temporal lobe of the pulse to the total pulse energy. In contrast to previous reports, post-compensation of the distorted spectral phase using dispersive mirrors enabled in-situ characterization of the ~7.6 fs-long near-UV pulses using self-diffraction frequency-resolved optical gating (SD-FROG) [12]. This direct measurement revealed that the quality of the output pulses reached values as high as $Q \sim 71\%$, which along with peak powers up to ~0.2 GW and excellent spatial mode quality, make this source very appealing for ultrafast spectroscopy.

## 2. Physical system

The proposed scheme relies on soliton-effect self-compression [10,13], followed by post-compensation of the chirp introduced along the optical path from the fiber output to the point of use. Upon propagation in ARR-PCF, pulses with sufficiently high peak power broaden spectrally due to self-phase modulation, generating a positive frequency chirp, which is compensated by the negative group velocity dispersion (GVD) of the hollow core [14]. If higher-order effects and Raman scattering are absent, the pulse would reach a temporal focus after propagating a distance $L_C \approx \sqrt{L_D L_{NL}/2}$ [15], $L_D = T_0^2/|\beta_2|$ being the dispersion length and $L_{NL} = 1/\gamma P_0$ the nonlinear length. Here, $T_0$ is the pump pulse duration, $P_0$ its peak power, $\beta_2$ the GVD and $\gamma$ the nonlinear coefficient of the fiber. Fig. 1(a) shows the compression length for bandwidth-limited pulses of 40 fs centered at 400 nm, as a function of fiber core diameter and filling gas pressure (properties which $\beta_2$ and $\gamma$ depend on). For a ~22 μm-core ARR-PCF such as that used in our experiments, reasonable compression lengths of the order of 1 m are obtained for gas pressures of ~250 mbar (indicated by the pink dots). Furthermore, since the self-compression quality is inversely proportional to the soliton order $N = \sqrt{L_D/L_{NL}}$, low $N$ values are required to achieve high enough $Q$, which again requires operation at low pressure (see inset of Fig. 1(a)). For our experimental parameters $N \sim 3$, permitting high-$Q$, 8-fold self-compression of the initial near-UV pulses, as we will see below.

Fig. 1(b) plots the wavelength dependence of the GVD for an ARR-PCF filled with Ar at 250 mbar (solid-blue line) and evacuated (solid-green line). The analytical calculation was based on approximating the propagation constant of the fundamental $LP_{01}$-like mode of the kagomé ARR-PCF using the expression derived for a wide-bore capillary [16]:

$$\beta_{01}(\omega, p, T) = k\sqrt{n_g^2(\omega, p, T) - \left(\frac{u_{01}}{ka}\right)^2}, \qquad (1)$$

where $k = \omega/c$ is the vacuum wavenumber, $n_g$ is the refractive index of the filling gas (at frequency $\omega$, pressure $p$ and temperature $T$), $a$ is the fiber core radius, and $u_{01} \approx 2.405$. Importantly, achieving anomalous dispersion in the UV (an essential ingredient for soliton-effect self-compression) requires the fiber core to be small and the gas pressure low. Moreover, the core wall must be extremely thin (below 100 nm) so as to push the first-order anti-crossing between the fundamental core mode and resonances in the thin glass walls into the deep UV, thereby ensuring broadband operation and shallow dispersion above 200 nm. There is currently a growing interest in another type of ARR-PCF with a simplified cladding composed of a single-ring of hollow tubes [17,18]. While these structures offer remarkably low attenuation [19,20] and polarization-maintaining properties [21] in the near-infrared and visible domains, the UV region remains a challenge [22] where kagomé-style fibers are still competitive.

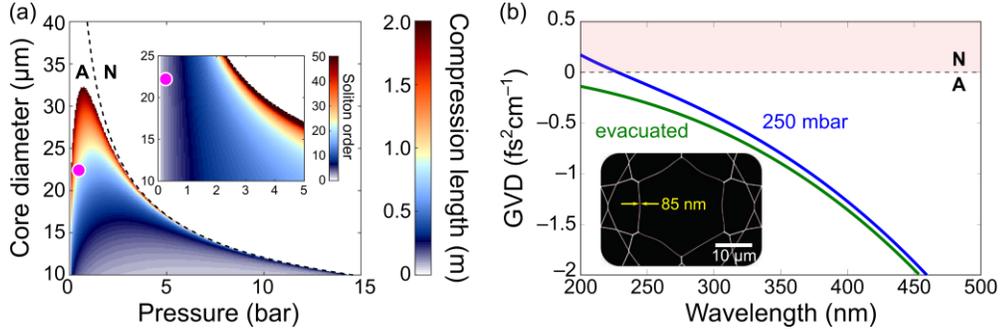

Fig. 1. (a) Compression length as a function of gas pressure and fiber core diameter. We consider bandwidth-limited, 40-fs-long pump pulses with 1.8 μJ centered at 400 nm. The dashed line marks the boundary between regions of normal (N) and anomalous (A) dispersion. Inset: Soliton order for a subset of parameters. The pink dot marks the operating point of the system, namely $N \sim 3$ and $L_c \sim 1.15$ m. (b) Group-velocity dispersion of the evacuated (solid-green) and Ar-filled (solid-blue) ARR-PCF as a function of wavelength. The zero-dispersion wavelength of the Ar-filled fiber is ~226 nm. Regions of normal (N) and anomalous (A) dispersion are highlighted by the light-red and white shading. Inset: Scanning electron micrograph of the fiber core structure.

## 3. Experimental set-up

The experimental set-up is sketched in Fig. 2(a). Energy-tunable pulses at 400 nm, obtained from a frequency-doubled Ti:sapphire amplified system, were spatially filtered and launched with ~70% efficiency into a 1.14 m-long kagomé-style ARR-PCF with ~22 μm-core diameter and ~85 nm average core-wall thickness (see inset in Fig. 1(b)). The fiber was enclosed between two gas cells hermetically sealed with $MgF_2$ windows, allowing selective gas pressurization. Eight bounces off a pair of negatively-chirped mirrors provided a total group delay dispersion (GDD) of ~ −400 $fs^2$, and a pair of thin fused silica wedges were used to finely balance the positive GDD introduced by the output $MgF_2$ window and the air-path to the SD-FROG set-up.

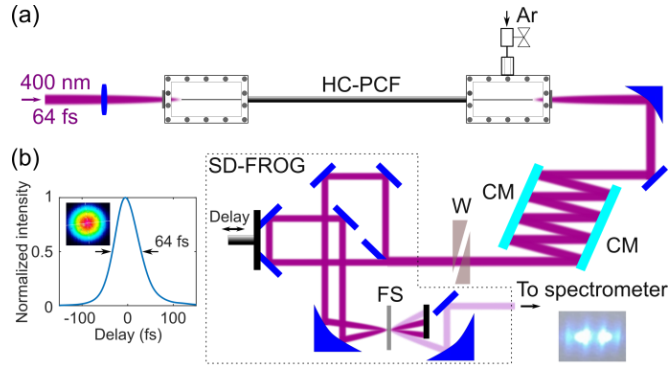

Fig. 2. (a) Experimental set-up. CM: Chirped mirrors; W: Wedges; FS: Fused silica plate. The dashed line encloses the home-built, all-reflective SD-FROG system. Inset: Self-diffracted orders on a view card placed at the position of the physical block. (b) Temporal profile of the pump pulses characterized using SD-FROG. Inset: Transverse beam profile of the pump pulses.

In the home-built, dispersion-free SD-FROG system (see Fig. 2(a)), a D-shaped mirror spatially divided the pulse into two identical replicas, the relative delay being controlled by a motorized stage. Upon interaction in a third-order nonlinear medium—in our case a thin fused-silica plate—the two pulses generated a thin transient grating capable of diffracting the same pulses in the Raman-Nath regime, resulting in the pattern depicted in the inset. The SD-FROG spectrogram was then recorded by spectrally resolving one of the first diffracted orders for a series of incremental delays. It is well-known that the limited sensitivity of the SD-FROG geometry requires sufficiently high input energy to generate a detectable signal in the first diffracted order. To meet this requirement, we positively chirped the ~40 fs bandwidth-limited pump pulses to ~64 fs duration in order to permit damage-free launching of microjoule levels of energy into the fiber, which provided strong enough output signal for complete temporal characterization using SD-FROG. The complete spatio-temporal profile of the pump pulses is shown in Fig. 2(b).

## 4. Results and discussion

When 3 µJ pulses were launched into the ARR-PCF, filled with Ar at 250 mbar, 2.4 µJ was measured at the fiber output. The spectrogram recorded by SD-FROG after self-compression of the input pulses is shown in Fig. 3(a). It is a very simple trace featuring a tilted central stripe that contains two main peaks and several satellite lobes. The slight negative slope of the SD-FROG spectrogram indicates that a small amount of negative GDD ~ −5 $fs^2$ from the chirped mirrors was not totally compensated by the thin wedges. An algorithm based on the extended ptychographic iterative engine [23] was then used to retrieve the pulses. The retrieved spectrogram (Fig. 3(b)) is in good agreement with the measured one, containing the same main features. Thus, the ptychographic algorithm yielded a pulse duration of ~7.6 fs (Fig. 3(c)) with a quality factor $Q \sim 71\%$ and excellent transverse beam quality. The main peak is surrounded by pre- and post-pulses, a clear signature of higher-order-soliton-effect self-compression [13]. Finally, the excellent agreement between the retrieved spectrum and an independent reference (Fig. 3(d)) further confirmed the reliability of the pulse retrieval procedure.

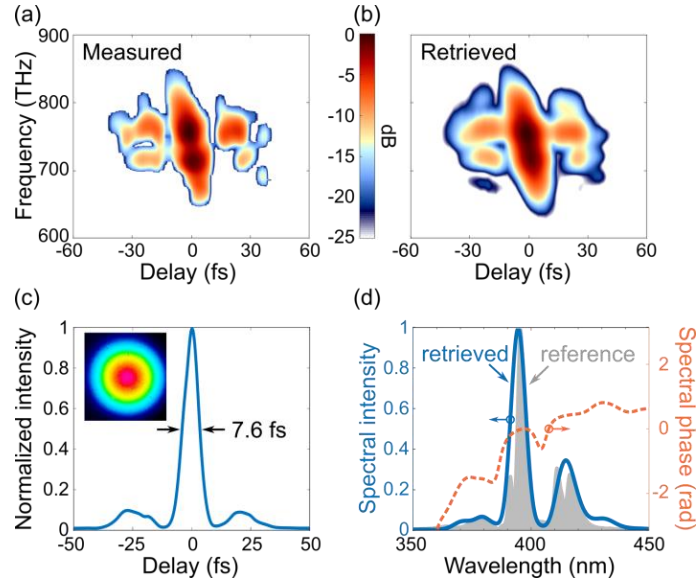

Fig. 3. (a) Measured and (b) retrieved SD-FROG spectrograms. (c) Retrieved normalized temporal intensity of the self-compressed pulses. Inset: Near-field transverse beam profile of the self-compressed pulses. (d) Retrieved normalized spectral intensity (solid-blue line), spectral phase (dashed-orange line), and measured reference spectrum (shaded gray).

To gain further insight into the near-UV self-compression process, we simulated the propagation dynamics of the measured pump pulses (Fig. 2(b)) along the fiber using the generalized nonlinear Schrödinger equation [11,14]. The simulated fiber characteristics were the same as in the experiment, the values of gas dispersion and nonlinearity being taken from [24] and [25]. The results of the simulations are displayed in Fig. 4. The dynamic interplay between nonlinear self-phase modulation and the negative fiber GVD causes the pump pulses to spectrally broaden (Fig. 4(a)) and temporally self-compress (Fig. 4(b)). Slight deviations from an initial quasi-Gaussian spectral shape (Fig. 5(a)), along with the presence of a small amount of third-order dispersion leads to an asymmetric, double-hump output spectrum, in good qualitative agreement with the experiments (Fig. 5(a)). Quantitatively, however, the bandwidth of the simulated spectrum is broader, extending up to 450 nm and supporting pulses with ~6.6 fs (solid-red line in Fig. 5(b)), slightly shorter than those measured by SD-FROG (shaded cyan in Fig. 5(b) for comparison).

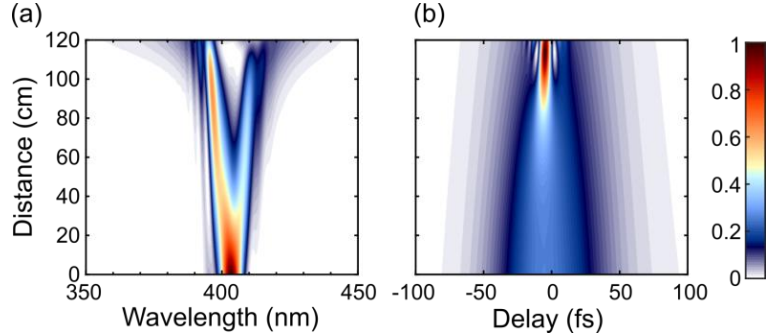

Fig. 4. Simulated spectral (a) and temporal (b) evolution of the real pump pulse propagating along the ARR-PCF employed in the experiment. To achieve the best agreement with the experimental results, the simulated input pulse energy was decreased by 40% (there is some uncertainty in the values of the nonlinear coefficients in the UV). The plots are normalized to the overall maximum intensity and the time delay is relative to a reference frame co-moving at the pump group velocity.

We attribute this discrepancy between simulated and measured pulse durations to the likely loss of some spectral components in the experiment along the pathway from the fiber output to the diagnostics. To support this assertion, we simulated this effect by numerically applying a super-Gaussian window to filter out the red tail of the simulated output spectrum (solid-red line in Fig. 5(a)) so that it would resemble more closely the experimental reference (shaded gray region in Fig. 5(a)). The spectrum obtained as a result of this short-pass filtering (dashed-purple line in Fig. 5(a)) indeed supports high-quality pulses of ~7.8 fs (dashed-purple line in Fig. 5(b)), in close agreement with the retrieval.

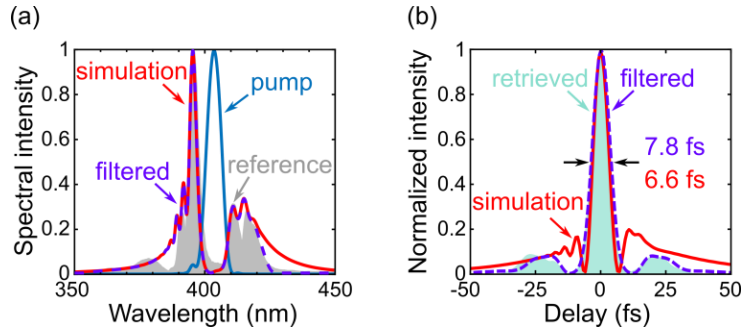

Fig. 5. (a) Simulated (solid-red line) and directly measured (shaded gray, see also Fig. 3(d)) spectral profiles of the out-coupled pulses, together with the initial pump spectrum (solid-blue line). The result of applying a super-Gaussian filter of 67 nm full-width at half-maximum and centered at 387 nm to the simulated spectrum is also displayed in dashed-purple line. (b) Simulated (solid-red line) and retrieved (shaded cyan) temporal profiles of the out-coupled pulses. For comparison, the temporal profile corresponding to the filtered spectrum shown in (a) is also included in dashed-purple line.

## 5. Conclusions and outlook

Near-UV, µJ-level pulses of tens of femtoseconds at 400 nm can be efficiently self-compressed down to few femtoseconds using a suitably designed meter-scale kagomé-style ARR-PCF filled with Ar at low pressure. By adding dispersion compensation elements after the fiber output, 8-fold compression of the input pulses to ~7.6 fs duration can be obtained, with very high spatio-temporal quality and peak powers as high as ~0.2 GW. Numerical modelling accurately describes the nonlinear soliton dynamics, confirming the spectro-temporal properties of the experimental self-compressed pulses. The system could be used "as is" for time-resolved applications in fields such as spectroscopy or materials science.

**Acknowledgments.** The work was funded by the Max Planck Society.

During the analysis of the results and preparation of the paper, David Novoa was supported by the European Regional Development Fund, the Spanish "Ministerio de Economía y Competitividad" under the project PGC2018-101997-B-I00, the Spanish "Ministerio de Ciencia e Innovación" under projects PID2021-122505OB-C31 and TED2021-129959B-C21 and the Gobierno Vasco/Eusko Jaurlaritza under projects IT1452-22 and ELKARTEK (KK 2021/00082 and KK 2021/00092).